# Formation of *c*BN nanocrystals by He$^+$ implantations of *h*BN

**R Machaka**[1,2,3]‡, **R M Erasmus**[1,2], **T E Derry**[1,2]

[1] DST/NRF Centre of Excellence in Strong Materials, University of the Witwatersrand, Private Bag 3, WITS 2050, RSA

[2] School of Physics, University of the Witwatersrand, Private Bag 3, WITS 2050, RSA

[3] iThemba LABS (Gauteng), Private Bag 11, WITS 2050, RSA

**Abstract.** The structural modifications of polycrystalline hexagonal boron nitride implanted with He$^+$ ion beams at energies between 200keV and 1.2MeV to fluences of $1.0 \times 10^{17}$ ions·cm$^{-2}$ were investigated using micro-Raman spectroscopy. The measured Raman spectra show evidence of implantation-induced structural transformations from the hexagonal phase to nanocrystalline cubic boron nitride, rhombohedral boron nitride and amorphous boron nitride phases. The first-order Longitudinal-Optical cBN phonon was observed to be downshifted and asymmetrically broadened and this was explained using the spatial correlation model coupled with the high ion implantation-induced defect density.

‡ Present address: School of Chemical and Metallugical Engineering, University of the Witwatersrand, Private Bag 3, WITS 2050, RSA.

## 1. Introduction

In recent years cubic boron nitride (cBN) has attracted considerable experimental and theoretical interest due to its rather unusual combination of superior physico-chemical properties and its enormous potential for technological applications. Several reviews have comprehensively discussed the structures, properties, and the significant technological importance of cBN, hexagonal boron nitride (hBN), and other boron nitride phases, see [1, 2] and references therein.

Although cBN does not occur in nature, it is the most stable phase with respect to other boron nitride phases under ambient conditions of temperature and pressure. However, a spontaneous hBN–to–cBN phase transformation is hindered by the very large activation energy barrier which exists between the respective phases. Bulk cBN is commercially synthesized by the high pressure and temperature induced hBN–to–cBN phase transformation in the presence of suitable solvent catalysts [3, 4]. Since its first successful synthesis in the mid-1950s, efforts have continued in searching for new routes that can be used to synthesize cBN under less extreme conditions of pressure and temperature.

It has been reported before, theoretically [4, 5] and experimentally [6, 7, 8], that under the non-equilibrium conditions of ion implantation, the sp$^2$ to sp$^3$ phase transformation can be carried out at normal pressure and moderate temperatures. The introduction of defects associated with ion implantation leads to dangling bonds between the basal layers of the graphitic materials and promote the subsequent corrugation of those planes [6]. Large concentrations of defects (vacancies/interstitials) present in graphite could lower the transformation barrier between diamond and graphite and permit diamond nucleation. Attempts have also been made by other research groups to use femtosecond [9] and ultraviolet [10] laser irradiation of graphite-like BN to try and influence the hBN–to–cBN phase transformation at ambient pressure and moderate temperatures.

In this work we employ micro–Raman spectroscopy ($\mu$-RS) to demonstrate the synthesis of nanocrystals of the cBN phase by He$^+$ ion implantation of the hBN phase.

## 2. Experimental Details

Hot–pressed 99.9% polycrystalline hBN powder samples supplied by Goodfellow UK were used throughout this study.

The samples were implanted with He$^+$ particle beams of energies between 200 keV and 1.2 MeV to fluences in the range of $1.0\times10^{15}$ ions·cm$^{-2}$ to $1.0\times10^{17}$ ions·cm$^{-2}$. The implantations were done using a iThemba LABS (Gauteng) modified 1.4MeV pressurized Cockcroft–Walton accelerator. All of the reported implantations were carried out at room temperature. The distributions of the radiation induced damage and the implant depths were calculated using the SRIM2008 program [11].

Micro–RS measurements were used to characterize the samples before and after implantation. The 514.5 nm line of an $Ar^+$ ion laser was used as an excitation source. The laser power was deliberately kept low to minimize temperature variation effects. An 1800 grooves/mm grating in the single spectrograph mode of a Jobin–Yvon T64000 Raman spectrometer was used. The $\mu$-RS measurements were performed using a 20× Olympus objective (the spot size diameter ∼1.5 $\mu$m). The Raman measurements were carried out at the Raman and Luminescence Laboratory in the School of Physics, University of the Witwatersrand.

## 3. Results and discussions

Figure 1 shows the $\mu$-RS spectrum measured on an unimplanted sample of $h$BN. The measured $\mu$-RS spectra exhibit an intense high–frequency $E_{2g}$ phonon mode that is characteristic of the B–N bond stretching vibrational mode within the $h$BN basal planes appearing around 1366cm$^{-1}$ [12, 13, 14, 15].

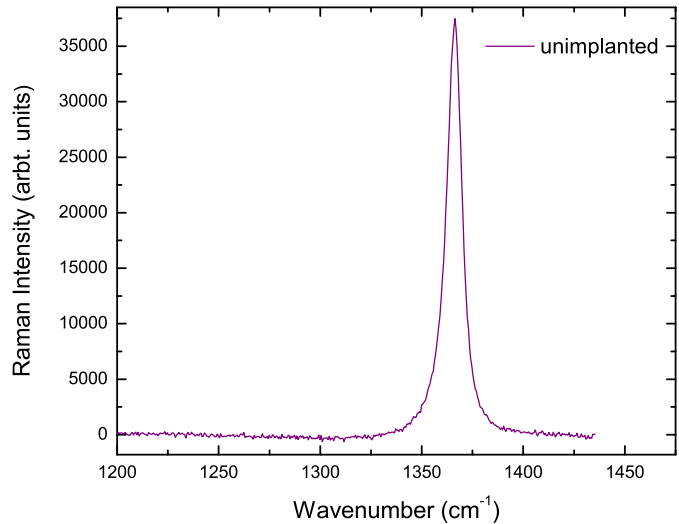

**Figure 1.** The $\mu$-Raman spectrum of an unimplanted hBN sample.

The $\mu$-Raman spectra measured on $h$BN samples that were implanted with 1.2MeV $He^+$ ions to fluences of $1.0 \times 10^{15}$ ions/cm$^2$ and $2.0 \times 10^{16}$ ions/cm$^2$ are presented in figure 2a and 2b, respectively. The spectra were measured at arbitrary points in the samples' implanted regions as well as the unimplanted regions. In the implanted region, the intensity of the $h$BN Raman peak greatly diminishes.

The $\mu$-Raman spectra measured on the 1.2MeV $He^+$ implanted $h$BN samples exhibit two broad Raman features in addition to the intense 1366cm$^{-1}$ $h$BN peak observed in the unimplanted $h$BN $\mu$-Raman measurement in figure 1: Peak 1, appearing around 752cm$^{-1}$ and 760cm$^{-1}$ for implantations at fluencies of up to $1.0 \times 10^{15}$ $He^+$ ions/cm$^2$ and $2.0 \times 10^{16}$ $He^+$ ions/cm$^2$, respectively; and Peak 2, a much weaker Raman feature, observed around 1290cm$^{-1}$ for implantation fluencies of up to $2.0 \times 10^{16}$ He$^+$ ions/cm$^2$, in figure 2b. Peak 2 is not evident in figure 2a.

We have attributed the origins of peak 1 to either crystalline/amorphous rhombohedral boron clusters [14], ion implantation-induced $h$BN–to–$r$BN phase transformation [16], or both [3]. $r$BN is a metastable sp$^2$ rhombohedral boron nitride allotrope.

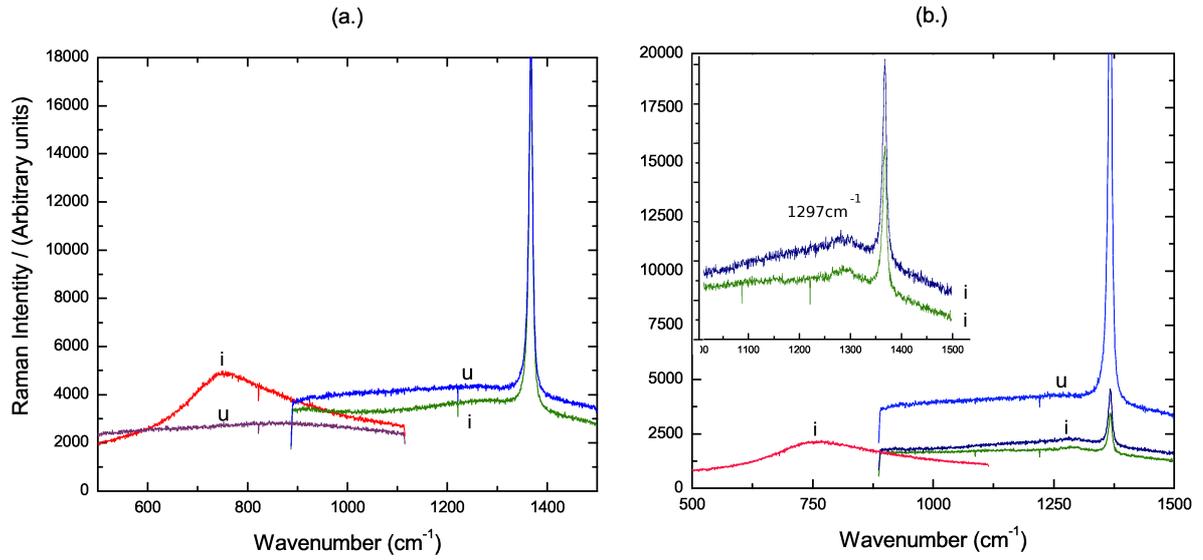

**Figure 2.** The $\mu$-Raman spectra taken at different points on the sample that was implanted to a fluence of [a.] $1.0 \times 10^{15}$ ions/cm$^2$ and [b] $2.0 \times 10^{16}$ ions/cm$^2$. The 752cm$^{-1}$ Raman feature is more dominant in [a], whilst the 1290cm$^{-1}$ feature is becoming evident in [b.]. The spectra marked $u$ indicate that the spectra were measured in the unimplanted region of the sample whilst those marked $i$ indicate that the spectra were measured in the implanted region. The insert in [b.] is also shown in figure 3a.

To further investigate the nature of peak 2, the results shown in figure 3 are $\mu$-RS spectra measured on $h$BN samples that were implanted to a fluence of $2.0 \times 10^{16}$ He$^+$ ions·cm$^{-2}$ at energies of (a) 1.2MeV, (b) 750keV, (c) 350keV, and at (d) 200keV. The intensity scales of the four spectra are arbitrary.

The spectrum of single-crystal $c$BN is characterized by two strong and distinctively separated Transverse-Optical (TO) and Longitudinal-Optical (LO) vibrational modes at 1056cm$^{-1}$ and 1305cm$^{-1}$, respectively [2, 3, 15]. However, the LO and TO vibrational modes are dependent on the size of the grains and the defect density or even the colour of the crystal. Hence the measured $\mu$-Raman spectra of ion implanted $h$BN would be expected to differ considerably from that of single-crystal $c$BN.

It has been demonstrated before [2, 3, 12] that in $c$BN films the Raman peaks downshift to lower wavenumbers and become substantially asymmetrically broadened, with increasing defect states and/or decreasing crystal sizes. Parayanthal and Pollak have developed a phonon confinement model to explain such changes in the $\mu$-RS spectra of micro- and nano-crystals [12, 17, 18]. Arora *et al.* [19] comprehensively reviews the Raman spectroscopy of optical phonon confinement in nanostructured materials.

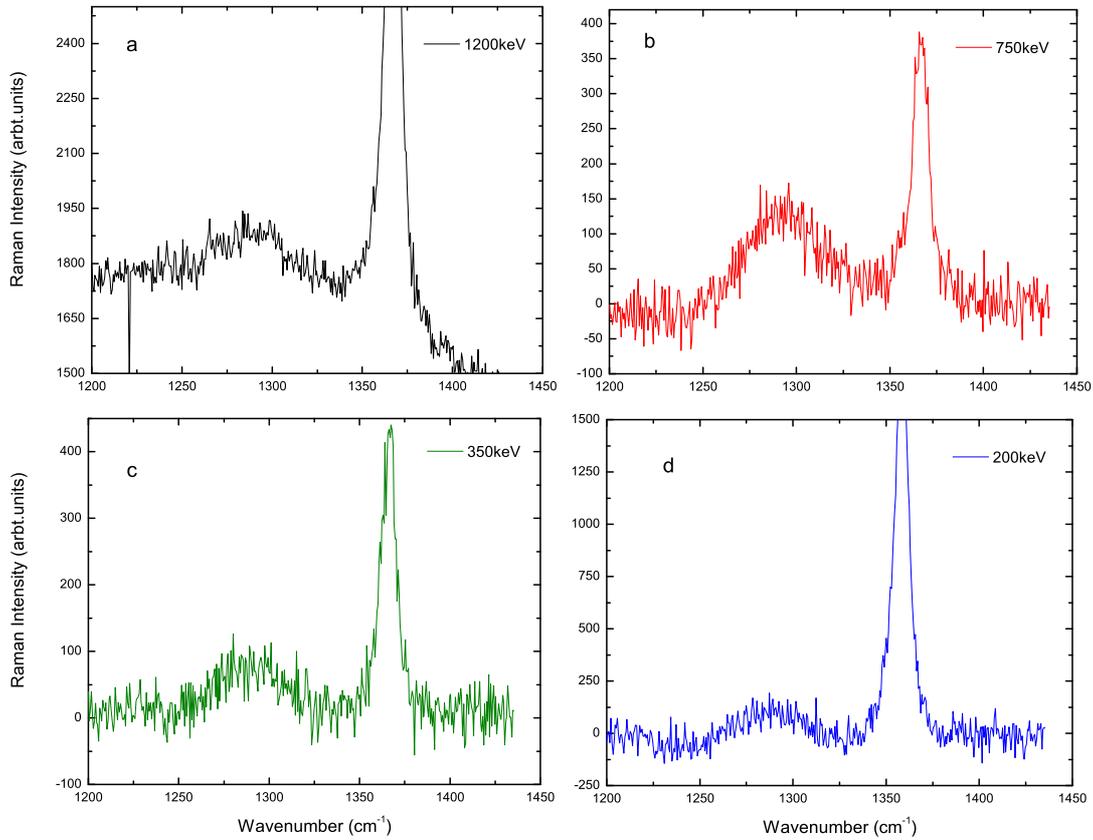

**Figure 3.** The $\mu$–Raman spectra of hBN samples implanted to fluences of $2.0\times10^{16}$ He⁺ ions·cm$^{-2}$ at energies of (a) 1.2MeV, (b) 750keV, (c) 350keV, and at (d) 200keV. This figure highlights the presence of peak 2 at $\sim1290$cm$^{-1}$ as a function of implantation energy. The vertical scales are arbitrary.

Similar downshifted, weak, and asymmetrically broadened Raman features to those in figure 3 have been reported before by other researchers for micro- and nano-crystals of cBN [12, 13], diamond [6, 15, 20], Si and Ge, and other group III-V semiconductor compounds [17, 19]. The wavenumbers of the broad features observed in our measurements (peak 2) correspond very well to the values reported in references [12] and [21]. We tentatively attribute the observed degradation of LO Raman features, with respect to that of the single crystal cBN, to the phonon confinement effect as a result of the existence of nanocrystals of cBN.

By curve fitting the observed Raman peaks to Lorentzian and Gaussian lineshapes,

values of the peak positions and intensities were obtained and normalized. For each implantation energy, the intensity of peak 2 was normalized to the respective principal hBN Raman peak that was measured outside the implanted region. From these values, two graphs were plotted; figure **??**a showing the variation of the normalized intensities of the Gaussian peaks ($I_{peak2}/I_{hBN}$) with the implantation energy and figure **??**b giving the variation of the position of peak 2 with the implantation energy.

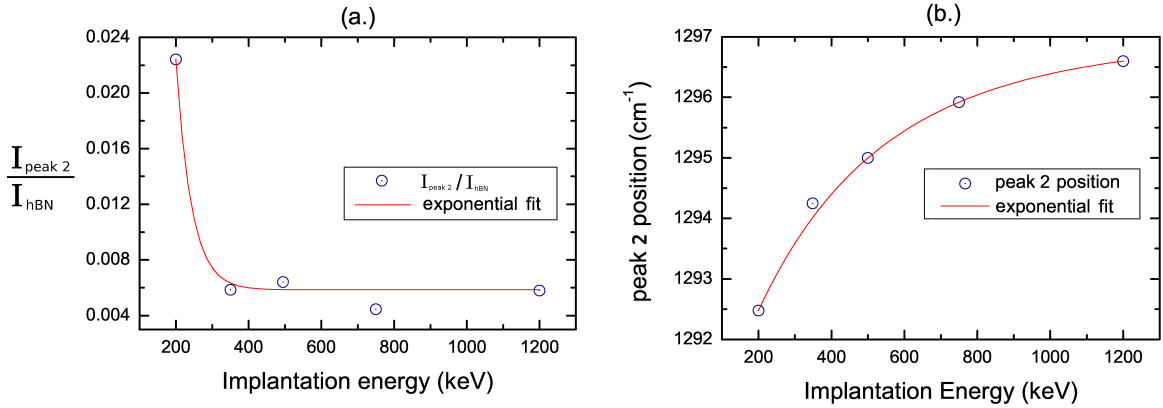

**Figure 4.** The dependence of (a.) the normalized intensities of the Gaussian peaks and (b.) the peak position of the Gaussian peaks with ion implantation energy.

The ratio $I_{peak2}/I_{hBN}$ can be linked to the change in the hybridization bonding from sp$^2$ to sp$^3$ (formation of $c$BN nanocrystals) and the position and shape of the Raman peak can indicate the crystallinity and order of the material. Figure **??**a shows an exponential growth in sp$^3$ hybridized states towards the lower implantation energy range. On the other hand, the position of peak 2 appears to be approaching a saturation as the implantation energy increases, possibly towards 1300cm$^{-1}$ in figure **??**b.

SRIM calculations of the inelastic damage profiles and the He$^+$ implant distributions as a function of depth were carried out in order to explain the rapid growth of sp$^3$ hybridized states towards the surface layers that is observed towards the lower implantation energy regime. It is noted that much of the damage profile created at 200keV lies within the penetration depth region (typically 1$\mu$m) of the probing Raman laser beam. It is also noted that the damage profile created at 1.2MeV lies mostly beyond this penetration depth region, however, the profile exhibits a long tail that overlaps this penetration depth region of the probing laser radiation. In the surface region that is being probed by the Raman laser, it is also possible that the nanocrystals formed at the lower energies are more disordered in structure or that the formed crystallites do not have any particular preferential orientation.

A critical dose of the order of 2.0×10$^{16}$ He$^+$ ions·cm$^{-2}$ was observed to favour the hBN–to–$c$BN phase transformation; He$^+$ implantations below the critical dose, of the order of 10$^{15}$ He$^+$ ions·cm$^{-2}$, were observed not to have influenced the hBN–to–$c$BN phase transformation in any observable way (see figure 2) whilst implantations at doses above the critical, of the order of $10^{17}$ He$^+$ ions·cm$^{-2}$ were observed only to produce extensive radiation damage of the sample as shown by the collapse of the principal *h*BN line at 1366cm$^{-1}$.

## 4. Summary and conclusions

The Raman characteristics of *c*BN do not usually emerge when the *c*BN is nanocrystalline and/or highly defective. We have presented experimental results showing the possibilities of using ion implantation of *h*BN to influence a structural transformation at room temperature that results in the formation of nanocrystals of the *c*BN phase. We have also shown that the measured *μ*-Raman spectra can be explained using the spatial correlation model that justifies the deviation of the observed spectra from that of *c*BN single crystals. Results tentatively suggesting that implantations at higher energies irrespective of the implantation dose might also influence an *h*BN–to–*r*BN phase transformation were also put forward.